\documentclass[twocolumn,aps,showpacs,prb,tightenlines,amsmath,amssymb]{revtex4}
\usepackage{graphicx}
\usepackage{amssymb}
\usepackage{dcolumn}
\usepackage{amsmath}
\usepackage{bm}

\usepackage{colordvi}
\newcommand{\bgreek}[1]{\mbox{\boldmath$#1$\unboldmath}}

\begin{document}

\title{Infinite spin diffusion length of any spin polarization  along 
    direction perpendicular to effective magnetic field from
Dresselhaus and Rashba spin-orbit couplings with identical strengths
 in (001) GaAs quantum wells}

\author{K.\ Shen}
\affiliation{Hefei National Laboratory for Physical Sciences at
Microscale and Department of Physics,
University of Science and Technology of China, Hefei,
Anhui, 230026, China}
\author{M.\ W.\ Wu}
\thanks{Author to  whom correspondence should be addressed}
\email{mwwu@ustc.edu.cn.}
\affiliation{Hefei National Laboratory for Physical Sciences at
Microscale and Department of Physics,
University of Science and Technology of China, Hefei,
Anhui, 230026, China}
\date{\today}
 
\begin{abstract}
In this note, we show that the latest spin grating measurement of
spin helix by Koralek {\em et al.} [Nature {\bf 458}, 610 (2009)] 
provides strong evidence of the infinite spin diffusion length
of any spin polarization along the 
direction perpendicular to the effective magnetic field  from the
 Dresselhaus and Rashba spin-orbit couplings with identical strengths 
in $(001)$ GaAs quantum wells,
predicted by Cheng {\em et al.} [Phys. Rev. B {\bf 75}, 205328 (2007)].

\end{abstract}


\maketitle
As one of the spintronic focuses, manipulations of electron spin in
 spin-orbit-coupled
solid state systems have been extensively investigated
due to the potential application in future quantum computation and quantum
information processing.\cite{spintronics} In
confined semiconductors with
zinc-blende structure, both the Dresselhaus
spin-orbit coupling (SOC) \cite{dress} due to the bulk inversion
asymmetry  and the Rashba one\cite{rashba} due to the structure
inversion asymmetry  are present, and the interplay of them gives rise
to amazing effects as reported in the
literature.\cite{averkiev,stan,wu1,wu2,wu3,bern,weber,schliemann,badalyan,koralek} For
example, an infinite spin relaxation time for
spin-polarization along the $(110)$ [or $(\bar 110)$ depending on the
sign of the Rashba term] direction
was predicted by Averkiev and Golub\cite{averkiev} in $(001)$ GaAs 
quantum wells (QWs) when the cubic term of the Dresselhaus SOC is excluded,
and the two leading terms, i.e., 
the linear Dresselhaus term and the Rashba one, are of the same strengths.
With the cubic term, the spin relaxation time along the $(110)$ direction 
becomes finite, but still much longer than those in the 
other directions.\cite{wu3}
Bernevig {\em et al.}\cite{bern} predicted a persistent spin helix
(PSH) by analyzing
the symmetry with identical Dresselhaus and Rashba strengths, which was
observed by Koralek {\em et al.}\cite{koralek} in the latest transient
spin grating experiment.\cite{tsg} 
Moreover, Cheng {\em et al.}\cite{wu1} studied the anisotropic
spin diffusion/transport in $(001)$ GaAs QWs with identical Dresselhaus and Rashba
strengths and predicted that in the direction perpendicular to the
effective magnetic field from the Dresselhaus and Rashba terms, 
i.e.,  $(\bar110)$ if the later is along the $(110)$ direction,\cite{note110} 
the spin diffusion length
is  infinite {\em regardless of} the spin polarization direction.
It seems rather strange that the spin diffusion length can be still
infinite for the spin polarization direction other than the direction of the
effective magnetic field.
According to the well-known relation $L_s=\sqrt{D_s\tau_s}$ widely
used in the transient spin grating works,\cite{defls} only a
finite spin diffusion length $L_s$ can be obtained for a spin
polarization other than the $(110)$ direction, because the spin diffusion coefficient $D_s$ and 
the spin relaxation time due to the D'yakonov-Perel' (DP)
mechanism\cite{dp} $\tau_s$ are both finite. However, to determine
the spin diffusion length via the transient spin
grating signal, as pointed out by Weng {\em et al.},\cite{wu2} the correct expression
should be
\begin{equation}
L_s=\sqrt[^4\!]{D_s^2\tau_{s1}\tau_s}/{\rm sin \tfrac{\phi}{2}}.
\label{eq1}
\end{equation}
This formula naturally gives an infinite spin diffusion
length because $\phi=0$ for identical Dresselhaus and Rashba
strengths.\cite{wu2}
 In the following part, we will discuss the underlying physics of the ideal spin
injection along the $(\bar110)$ direction and show that the corresponding injection mode
is the same as the PSH mode\cite{bern} in the transient spin
grating along the $(\bar 110)$ direction.\cite{note110} Therefore, the
observation of the PSH also gives
strong evidence of the ideal spin injection/diffusion efficiency along
the $(\bar110)$ direction
predicted by Cheng {\em et al.}.\cite{koralek}
Similar to the spin relaxation time along the $(110)$ direction, the small cubic
term of the Dresselhaus SOC makes the spin diffusion length 
along the $(\bar110)$ direction finite, but still much longer than those along
other directions.

The origin of the infinite spin diffusion length in the steady-state
spin injection along the $(\bar110)$ direction in $(001)$ QWs was explained by Cheng {\em
 et al.}, associated with the inhomogeneous 
broadening.\cite{wu1} For spin transport along the $x$-axis, the inhomogeneous broadening
is determined by the procession frequency ${\bgreek 
  \omega}_{\bf k}=\frac{m}{2\hbar^2k_x}{\bf \Omega}({\bf k})$,\cite{wu1,wu2,wu4,wu5}
with $\bf \Omega({\bf k})$ and $m$ representing the effective magnetic
field from the Dresselhaus and/or Rashba terms
with momentum $\bf k$ and the effective electron mass, respectively. When the
Rashba coefficient $\alpha$ equals to the 
linear Dresselhaus coefficient $\beta$, ${\bgreek \omega}_{\bf k}$
reads\cite{wu1} 
\begin{eqnarray}
\nonumber
{\bgreek \omega}_{\bf k}&=& \frac{m}{2\hbar^2}\Bigg\{ 2\beta\left[{\rm
        sin}\left(\theta-\frac{\pi}{4}\right)+ {\rm
        cos}\left(\theta-\frac{\pi}{4}\right)\frac{k_y}{k_x}\right]{\bf
    \hat n}_0  \\\nonumber
&&+\gamma\left(\frac{k_x^2-k_y^2}{2}{\rm sin}2\theta+k_xk_y{\rm
    cos}2\theta \right)\left(
\begin{array}{c}
\tfrac{k_y}{k_x}\\ -1\\ 0
\end{array}
\right)\Bigg\}\ , 
\end{eqnarray}
with $\hat {\bf n}_0$ denoting
the crystal direction $(110)$. $\theta$ stands for the angle between
the injection direction $x$ and the $(100)$ crystal axis.
 The second term with coefficient
$\gamma$ on the right side of the equation  is the cubic Dresselhaus
term, which we neglect in the
following discussion for simplification. 
 For spin injection along the
$(\bar110)$ direction, i.e., $\theta=3\pi/4$,
the precession frequency becomes ${\bgreek
  \omega}_{\bf k}=\tfrac{m\beta}{\hbar^2}\hat{\bf n}_0$, which is independent
of the momentum $\bf k$. Therefore
there is no inhomogeneous broadening in this case.\cite{wu1} As a result,
the spin polarization diffuses into the semiconductor without
any decay of the amplitude for the steady-state spin injection condition, even with
all the relevant spin-conserving 
scatterings, such as, the electron-impurity, electron-phonon and
electron-electron scatterings.
Moreover, the spin-$1/2$ electron ensemble with a unique precession frequency $\bgreek
\omega$ gives the spatial spin oscillation period
$L_0=2\pi/(2m\beta)$, along the diffusion length.\cite{wu1}

The infinite spin diffusion length was also associated with the
transient spin grating by Weng {\em et al.}.\cite{wu2}
In the transient spin grating along the $(\bar110)$ direction, the
initial spin-orientation wave is
composed of two spin helices, which decay with different
rates.\cite{bern,wu2,weber,koralek} For $\alpha=\beta$, the slow-decay
helix exactly matches the spin precession mode 
when the wave vector $q$ equals to $2m\beta$ (the corresponding spatial
period $L_0=2\pi/(2m\beta)$).\cite{note}
Therefore, this mode becomes PSH with the relaxation time $\tau_-=\infty$ and dominates
the steady-state spin injection,\cite{wu2} while the other one decays
with finite $\tau_+$
as usual.\cite{bern} By 
directly integrating the transient spin grating signal over the 
time from 0 to $\infty$ and the wave vector $q$ from $-\infty$ to
$\infty$, the steady-state
spin injection solution is $S_z(x)=S_z(0)e^{-x/L_s}{\rm
  cos}(x/L_0+\psi)$ with the spatial oscillation period
$L_0=\sqrt[^4\!]{D_s^2\tau_{s1}\tau_s}/{\rm cos \tfrac{\phi}{2}}$ and
the spin diffusion length $L_s$ [shown in Eq.\,(\ref{eq1})].\cite{wu2}
 Since $\tau_{1s}=\tau_s$ and $\phi=0$ for $\alpha=\beta$,\cite{wu2} one obtains
 $L_0=2\pi/(2m\beta)$ and $L_s=\infty$, which are coincident with those 
predicted by
 Cheng {\em et al.}.\cite{wu1}
Thus, the observation of the PSH strongly supports the
prediction of the infinite spin diffusion length by Cheng {\em et al.}.\cite{wu1}

In real systems, both the spin diffusion length in the
steady-state spin injection and the spin lifetime of the PSH mode in the
transient spin grating are finite
because of the cubic Dresselhaus term. With this
term, the optimal condition is $\alpha=\beta-\gamma k^2/4$ instead of
$\alpha=\beta$,\cite{koralek} as predicted theoretically.\cite{stan,wu1}. 
However, the direct quantitative 
measurement of the largest spin diffusion length 
still requires further experimental efforts.\cite{fabian}

This work was supported by the Natural Science Foundation of China
under Grant No.~10725417. We would like to thank
J. Fabian for helpful discussions.


\begin{thebibliography}{0}
\bibitem{spintronics} {\em Semiconductor Spintronics and Quantum
    Computation}, ed. by D. D. Awschalom, D. Loss, and N. Samarth
  (Springer-Verlag, Berlin, 2002); I. \u Zuti\'c, J. Fabian, and S. Das
  Sarma, Rev. Mod. Phys. {\bf 76}, 323 (2004).

\bibitem{dress} G. Dresselhaus, Phys. Rev. {\bf 100}, 580 (1955).

\bibitem{rashba} Y. A. Bychkov and E. I. Rashba, J. Phys. C {\bf 17},
  6039 (1984); Pis'ma Zh. Eksp. Teor. Fiz. {\bf 39}, 66 (1984) 
  [JETP Lett. {\bf 39}, 78 (1984)].

\bibitem{averkiev} N. S. Averkiev and L. E. Golub, Phys. Rev. B {\bf
    60}, 15582 (1999).
\bibitem{schliemann} J. Schliemann, J. C. Egues, and D. Loss,
  Phys. Rev. Lett. {\bf 90}, 146801 (2003).

\bibitem{wu3} J. L. Cheng and M. W. Wu, J. Appl. Phys. {\bf 99},
  083704 (2006).

\bibitem{stan} T. D. Stanescu and V. Galitski, Phys. Rev. B {\bf 75},
  125307 (2007).

\bibitem{bern} B. A. Bernevig, J. Orenstein, and S. C. Zhang,
  Phys. Rev. Lett. {\bf 97}, 236601 (2006).

\bibitem{weber} C. P. Weber, J. Orenstein, B. A. Bernevig,
  S. C. Zhang, J. Stephens, and D. D. Awschalom, Phys. Rev. Lett. {\bf
  98}, 076604 (2007).
\bibitem{badalyan} S. M. Badalyan, A. Matos-Abiague, G. Vignale, and
  J. Fabian,  Phys. Rev. B {\bf 79}, 205305 (2009).
\bibitem{koralek}J. D. Koralek, C. P. Weber, J. Orenstein, B. A. Bernevig,
    S. C. Zhang, S. Mack, and D. D. Awschalom, Nature {\bf 458}, 610
    (2009).
\bibitem{wu1} J. L. Cheng, M. W. Wu, and I. C. da Cunha Lima, Phys. Rev. B {\bf
    75}, 205328 (2007).
\bibitem{wu2} M. Q. Weng, M. W. Wu, and H. L. Cui, J. Appl. Phys. {\bf
  103}, 063714 (2008).
\bibitem{tsg} A. R. Cameron, P. Riblet, and A. Miller,
  Phys. Rev. Lett. {\bf 76}, 4793 (1996).
\bibitem{note110} It depends on the direction of the effective magnetic field. For
  example, the infinite spin diffusion length can be obtained along
  the $(110)$ direction when the effective magnetic field turns to
  the $(\bar110)$ direction by changing the
  sign of the Rashba SOC coefficient through the electrical approach.
\bibitem{defls} K. V. Kavokin, Phys. Status Solidi A {\bf 190}, 221
  (2002); C. P. Weber, N. Gedik, J. E. Moore, J. Orenstein,
  J. Stephens, and D. D. Awschalom, Nature {\bf 437}, 1330 (2005).
\bibitem{dp} M. I. D'yakonov and V. I. Perel',
  Zh. \'Eksp. Teor. Fiz. {\bf 60}, 1954 (1971) [Sov. Phys. JETP {\bf
    33}, 1053 (1971)]; Fiz. Tverd. Tela (Leningrad) {\bf 13}, 3581
  (1971) [Sov. Phys. Solid State {\bf 13}, 3023 (1972)].
\bibitem{wu4} M. Q. Weng and M. W. Wu, Phys. Rev. B {\bf 66}, 235109
  (2002); J. Appl. Phys. {\bf 93}, 410 (2003).
\bibitem{wu5} J. L. Cheng and M. W. Wu, J. Appl. Phys. {\bf  101},
  073702 (2007).
\bibitem{note} The different expression of $q$ compared to Ref.\,\onlinecite{bern} comes from
  the specific definition of the SOC coefficients.
\bibitem{fabian} J. Fabian, Nature {\bf 458}, 580 (2009).
\end{thebibliography}
\end{document}